# DYNAMIC FRONT TRANSITIONS AND SPIRAL-VORTEX NUCLEATION

Christian Elphick
*Department of Physics*
*Universidad Técnica Federico Santa Maria*
*Valparaiso, Casilla 110-V, Chile*

Aric Hagberg*
*Program in Applied Mathematics*
*University of Arizona*
*Tucson, AZ 85721, USA*

Ehud Meron
*The Jacob Blaustein Institute for Desert Research and The Physics Department*
*Ben-Gurion University*
*Sede Boker Campus 84990, Israel*

(December 1994)

This is a study of front dynamics in reaction diffusion systems near Nonequilibrium Ising-Bloch bifurcations. We find that the relation between front velocity and perturbative factors, such as external fields and curvature, is typically multivalued. This unusual form allows small perturbations to induce dynamic transitions between counter-propagating fronts and nucleate spiral vortices. We use these findings to propose explanations for a few numerical and experimental observations including spiral breakup driven by advective fields, and spot splitting.

05.45.+b, 82.20Mj

## I. INTRODUCTION

Nonequilibrium Ising-Bloch (NIB) transitions [1] have been identified recently as important mechanisms of pattern formation in reaction-diffusion [2,3] and liquid crystal [4] systems. Mathematically, a NIB transition amounts to a pitchfork bifurcation where a stationary (Ising) front loses stability to a pair of counter-propagating (Bloch) fronts. The coexistence of two Bloch fronts far beyond the bifurcation allows the formation of regular patterns such as periodic traveling domains and rotating spiral waves.

In this paper we show that near a NIB bifurcation spontaneous transitions between the two Bloch fronts become feasible and that a number of apparently unrelated phenomena can be understood in terms of these transitions. Front transitions of this kind can be induced by extrinsic perturbations, such as advective fields, or intrinsic perturbations, such as front interactions and curvature. They may occur uniformly along the front, reversing its direction of propagation, or locally, nucleating spiral-vortex pairs. We study the mechanism of such transitions and suggest explanations for two experimental observations: spiral breakup induced by the onset of convection in chemical reactions [5], and spot splitting [6]. We suggest that these experiments, the transition from labyrinthine patterns to spiral turbulence found numerically in [7], and possibly other breakup phenomena [8,9] are all realizations of the same mechanism.

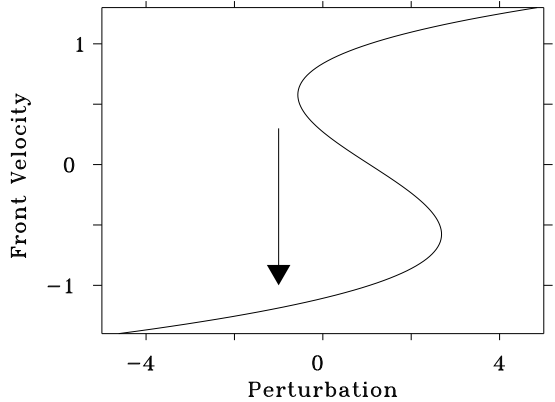

FIG. 1. A typical front-velocity vs. perturbation graph near a NIB bifurcation. The figure is a plot of $V$ vs. $I$ according to the stationary form of (13) with $\alpha_0 = -2.0$ and $a_1 = 5.0$.

The key feature underlying front transitions is a multivalued dependence of front velocity on extrinsic and intrinsic perturbations close to a NIB bifurcation. Fig. 1 shows a typical form of such a relation. The upper and lower branches correspond to the two counter-propagating Bloch fronts. At least one of them terminates at a *realizable* perturbation strength. A perturbation that drives the system to the end of a given Bloch front branch induces a transition to the other branch as indicated by the arrow.

We consider an activator-inhibitor reaction-diffusion system exhibiting a NIB bifurcation [3]:

$$u_t = u - u^3 - v + \nabla^2 u, \qquad (1a)$$

$$v_t = \epsilon(u - a_1 v - a_0) + \mathbf{J} \cdot \nabla v + \delta \nabla^2 v, \qquad (1b)$$

where $\mathbf{J}$ represents an advective vector field. In the context of chemical reactions $u$ and $v$ represent concentrations of two key chemical species, and $\mathbf{J}$ may stand for a convective flow field. We note that there is no a priori reason to assume that both species are advected differently by the flow field. However, chemical reactions normally involve many more than two chemical species. Two-variable models like (1) are obtained by adiabatic elimination of fast reacting species, and in this process transport terms are renormalized. As shown by Dawson et al. [10] this can lead to an effective differential flow as assumed in (1). The external field $\mathbf{J}$ may also stand for an electric field which advects ionic species [8,11].

In section 2 we use equations (1), simplified by setting $\delta = 0$, to derive an evolution equation for the front velocity near a NIB bifurcation. We obtain a multivalued relation between the front velocity and the advective perturbation $\mathbf{J}$, and show that transitions between the different front velocity branches follow from the gradient nature of the evolution equation. In section 3 we study the relation between the front velocity and the advective perturbation for $\delta/\epsilon \gg 1$ and at arbitrary distance from the NIB bifurcation. The multivalued nature of this relation near the NIB bifurcation is used to demonstrate how spiral breakup can be induced by an advective field. The relation to convection induced disorder in the Belousov-Zhabotinsky (BZ) reaction [5] is discussed. In section 4 we show that multivalued relations between the front velocity and intrinsic curvature perturbations near a NIB bifurcation can be used to explain spot splitting [6] and transitions from labyrinthine patterns to spiral turbulence [7]. We conclude in section 5 with a discussion of further possible implications of this work.

## II. THE DYNAMICS OF FRONT TRANSITIONS

To study front transitions we consider the simpler case of a non-diffusive $v$ field, $\delta = 0$, in one space dimension, $x$, and assume first a symmetric system, $a_0 = 0$, with no external field, $\mathbf{J} = 0$. We choose $a_1$ such that equations (1) describe a bistable medium having two linearly stable, stationary homogeneous states: an up state, $(u_+, v_+)$, and a down state, $(u_-, v_-)$, where $u_+ = -u_- = \sqrt{(1 - a_1^{-1})}$, and $v_+ = -v_- = a_1^{-1} u_+$. In Ref. [3] we studied one-dimensional front solutions of (1) propagating at constant speeds. The following is a summary of the main findings. The system (1) has a stationary Ising front solution,

$$u_0(x) = -u_+ \tanh(u_+ x/\sqrt{2}), \quad v_0(x) = a_1^{-1} u_0(x), \quad (2)$$

connecting the up state at $x = -\infty$ to the down state at $x = \infty$. This front solution exists for all $\epsilon > 0$. For $\epsilon < \epsilon_c = a_1^{-2}$ two additional Bloch front solutions appear, propagating at velocities

$$c = \pm\left[\frac{5}{2u_+^2}(\epsilon_c - \epsilon)\right]^{1/2}. \qquad (3)$$

They correspond to an up state invading the down state (positive speed) and to a down state invading the up state (negative speed). Their leading-order forms for $|c| \ll 1$ are

$$u(x,t) = u_0(x - ct), \quad v(x,t) = a_1^{-1} u_0(x - ct + ca_1), \quad (4)$$

[3]. To that order, the Bloch front structure differs from the Ising front structure in that the $v$ field is translated with respect to the $u$ field by an amount proportional to $c$.

Our objective is to derive an evolution equation for the front velocity, $C(t)$, near the NIB bifurcation, whose solutions describe dynamic transitions between the two Bloch fronts. We introduce a normalized front velocity, $V(t)$, such that $V = 0$ corresponds to the Ising front and $V = \pm 1$ to the Bloch fronts. The actual front velocity is given by $C = cV$, where $c$ is a positive constant to be identified with the absolute value of the Bloch front velocity given by (3). We derive an evolution equation for $V$ using asymptotic expansions in the small positive parameter $c$ and demanding uniform validity. We write

$$u(z,T) = u_0(z) + \sum_{n=1}^{\infty} c^n u_n(z,T), \qquad (5a)$$

$$v(z,T) = v_0(z) + \sum_{n=1}^{\infty} c^n v_n(z,T), \qquad (5b)$$

$$\epsilon = \sum_{n=0}^{\infty} c^n \epsilon_n, \qquad (5c)$$

where $z = x - \zeta(T)$ and $T = c^2 t$. In terms of the front position, $\zeta$, the normalized velocity is given by $V(T) = c^{-1}\zeta_t = c\zeta_T$. We use these expansions in (1) and solve for the corrections, $u_1, u_2, ...$ and $v_1, v_2, ...$, to the fields. At order $c$ we obtain

$$\mathcal{L}\begin{pmatrix} u_1 \\ v_1 \end{pmatrix} = -V\begin{pmatrix} u_{0z} \\ v_{0z} \end{pmatrix}, \qquad (6)$$

where

$$\mathcal{L} = \begin{pmatrix} \partial_{zz} + 1 - 3u_0^2 & -1 \\ \epsilon_0 & -\epsilon_0 a_1 \end{pmatrix}. \qquad (7)$$

Since the adjoint operator, $\mathcal{L}^\dagger$, has the null vector $(u_{0z}, -\epsilon_0^{-1} v_{0z})^T$, a solvability condition leads to $\epsilon_0 = \epsilon_c = a_1^{-2}$. Solving (6) we find

$$u_1 = 0, \qquad v_1 = V u_{0z}. \qquad (8)$$

At order $c^2$ we obtain

$$\mathcal{L} \begin{pmatrix} u_2 \\ v_2 \end{pmatrix} = \begin{pmatrix} 0 \\ -\epsilon_1 a_1 v_1 - V v_{1z} \end{pmatrix}. \qquad (9)$$

Solvability of (9) yields $\epsilon_1 = 0$ and the solutions

$$u_2 = \frac{1}{2} a_1 V^2 z u_{0z}, \quad v_2 = \frac{1}{2} V^2 z u_{0z} + a_1 V^2 u_{0zz}. \qquad (10)$$

Finally, at order $c^3$

$$\mathcal{L} \begin{pmatrix} u_3 \\ v_3 \end{pmatrix} = \begin{pmatrix} -V u_{2z} \\ v_{1T} - V v_{2z} + a_1 \epsilon_2 v_1 \end{pmatrix}. \qquad (11)$$

Solvability of (11) gives the evolution equation for the front velocity;

$$V_T = \mu V (1 - V^2), \qquad \epsilon < \epsilon_c, \qquad (12)$$

where $\mu = a_1 |\epsilon_2|$ and $\epsilon_2 = \int_{-\infty}^\infty u_{0z} u_{0zzz} dz / \int_{-\infty}^\infty u_{0z}^2 = -2 u_+^2 / 5$. Equation (12) reproduces the Ising and the Bloch front solutions of (1) as constant speed solutions with $V = 0$ and $V = \pm 1$, respectively. Since $\epsilon = \epsilon_c - |\epsilon_2| c^2 + \mathcal{O}(c^4)$, the Bloch front velocities, $C = \pm c$, coincide with (3). Equation (12) also contains information about the stability of these solutions (for $\epsilon < \epsilon_c$) with respect to Galilean boosts; the Ising front is unstable and the Bloch fronts are stable.

Next, we consider the nonsymmetric case, $a_0 \neq 0$, with constant advective perturbation $\mathbf{J} = J \hat{\mathbf{x}}$. For simplicity we take $a_0 = c^3 \alpha_0$ and $J = c^3 I$, where $\alpha_0$ and $I$ are of order unity. The order parameter equation now reads

$$V_T = \mu V (1 - V^2) + \nu \alpha_0 + \frac{1}{a_1} I, \qquad (13)$$

where $\nu = \frac{3}{\sqrt{2} a_1 (a_1 - 1)}$. One immediate result from (13) is a multivalued relation, of the form shown in Fig. 1, between the constant velocity of a front and the advective perturbation $I$ (or $J$). Additionally, equation (13) describes a gradient flow (despite the nongradient nature of (1)) derivable from the potential

$$\mathcal{U} = \mu \frac{V^2}{2} \left[ \frac{V^2}{2} - 1 \right] - \left( \nu \alpha_0 + \frac{1}{a_1} I \right) V.$$

In a range of $I$ values, $I_{min}(a_0) < I < I_{max}(a_0)$, $\mathcal{U}(V)$ is a double well potential. The wells at $V = V_- < 0$ and $V = V_+ > 0$ pertain to Bloch fronts propagating to the left and to the right, respectively. Decreasing $I$ below $I_{min}$ causes the well at $V_+$, or the Bloch front that propagates to the right, to vanish. Since the dynamics are gradient, convergence to the well at $V = V_-$ follows. This amounts to a transition from a Bloch front propagating to the right (upper branch in Fig. 1) to a Bloch front propagating to the left (lower branch in Fig. 1), induced by the advective perturbation $J$.

We have considered the case $\delta = 0$, which is simpler for analysis, and showed that a perturbation driving a given Bloch front branch past its end point induces a transition to the other Bloch front branch. In the following sections we consider the more general case, $\delta \neq 0$, by studying constant speed front solutions for parameters satisfying $\delta / \epsilon \gg 1$. The analysis culminates in velocity - perturbation relations valid both near and far from the NIB bifurcation, but does not contain the information about the dynamics of front transitions derived in this section. Numerical results, however, indicate that the same dynamical behavior holds for $\delta \neq 0$ as well; a front transition is induced when a given front reaches the end point of a velocity branch.

## III. SPIRAL BREAKUP INDUCED BY AN ADVECTIVE FIELD

The earliest observations of spiral breakup in chemical reactions occurred in open cells containing the BZ reagents [5]. Convection in the form of Bénard cells, induced by evaporative cooling of the free surface, initiated processes of spiral breakup and spiral wave nucleation that destroyed the order of the pre-existing chemical pattern. We do not attempt a complete description of this system. Instead we use a simplified model, including what we believe are the essential ingredients responsible for this behavior (see discussion at the end of the section). We consider equations (1) in two space dimensions near the NIB bifurcation, assume $\delta / \epsilon \gg 1$, a condition normally met in experiments on the BZ reaction, and take a time independent advective field with hexagonal spatial structure (to model the convective Bénard cells).

Before simulating equations (1) with an hexagonal advective field, we derive the relation between the front velocity and a constant advective field $J$ for $\delta / \epsilon \gg 1$. This relation will help us to identify conditions leading to spiral breakup. In Refs. [3,12] we studied the NIB bifurcation for $\delta / \epsilon \gg 1$ in the absence of an advective field ($\mathbf{J} = 0$), using a singular perturbation approach (see also Ref. [13]). In this parameter range the spatial variation of the $v$ field is on a scale much longer than that of $u$. In the narrow front region where the variation of $u$ is of order unity, $v$ is approximately constant. Solving (1) (with $\mathbf{J} = 0$) in this inner region yields the relation

$$C_0 = -\frac{3}{\sqrt{2}} v_f,$$

between the front velocity, $C_0$, and the (small) constant value, $v_f$, of the $v$ field at the front position. Away from the narrow front region the $u$ field is enslaved to the $v$ field, $u = u_\pm(v)$, where $u_\pm(v)$ solve $u^3 - u - v = 0$. Solving (1) (with $\mathbf{J} = 0$) in the outer regions on both sides of the front and matching the solutions at the front position yield another relation between $C_0$ and $v_f$,

$$v_f = -\frac{C_0}{q^2(C_0^2 + 4\eta^2 q^2)^{1/2}} - \frac{a_0}{q^2}.$$

Eliminating $v_f$ we get the following implicit relation for the front velocity [3,12]:

$$C_0 = F(C_0, \eta),$$
$$F(X, Y) = \frac{3X}{\sqrt{2}q^2(X^2 + 4q^2Y^2)^{1/2}} + C_\infty, \qquad (14)$$

where $\eta^2 = \epsilon\delta$, $q^2 = a_1 + 1/2$, and $C_\infty = \frac{3a_0}{\sqrt{2}q^2}$. A graph of $C_0$ vs. $\eta$ (or $C_0$ vs. $\epsilon$ at constant $\delta$) yields a NIB bifurcation diagram valid at any distance from the bifurcation point as long as $\delta/\epsilon \gg 1$.

Equations (14) can be used to find a relation for the front velocity $C$ in the presence of a constant perturbation $J$. Consider a planar front solution of (1), with $\mathbf{J} = J\hat{\mathbf{x}}$, propagating at a constant velocity $C$ in the $x$ direction. It satisfies the equations

$$u'' + Cu' + u - u^3 - v = 0, \qquad (15a)$$

$$\delta v'' + (C+J)v' + \epsilon(u - a_1 v - a_0) = 0, \qquad (15b)$$

where the prime denotes differentiation with respect to the single independent variable $x - Ct$. Let us now rewrite equations (15) in the form

$$u'' + Cu' + u - u^3 - v = 0, \qquad (16a)$$

$$\delta_J v'' + Cv' + \epsilon_J(u - a_1 v - a_0) = 0, \qquad (16b)$$

where $\epsilon_J = \epsilon\Delta_J$, $\delta_J = \delta\Delta_J$, and $\Delta_J = \frac{C}{C+J}$. According to (16), front solutions of (15), the perturbed system with parameters $\epsilon$ and $\delta$, are equivalent to front solutions of an unperturbed system with effective parameters $\epsilon_J$ and $\delta_J$. Using relation (14), obtained for the unperturbed system, and replacing $C_0$, $\epsilon$, and $\delta$ by $C$, $\epsilon_J$, and $\delta_J$, respectively, we find

$$C = F(C+J, \eta), \qquad (17)$$

where $F$ is given by (14). Figs. $2a - 2c$ show graphs of $C$ vs. $J$, obtained by solving (17) for $C$ deep in the Bloch regime ($2a$), near the NIB bifurcation ($2b$), and deep in the Ising regime ($2c$). The single velocity branch deep in the Ising regime folds into three connected branches near the NIB bifurcation with a multivalued form resembling the $V$ vs. $I$ graph obtained from (13) for $\delta = 0$. Deep in the Bloch regime the three branches become practically

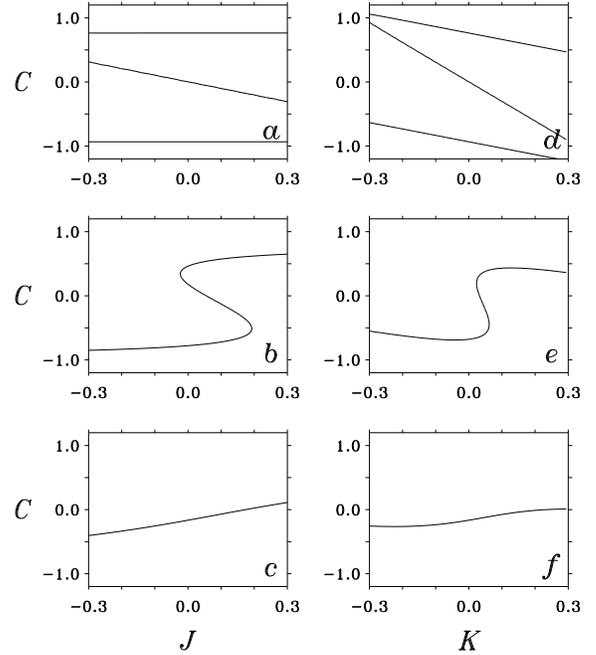

FIG. 2. Front velocity, $C$, vs. advective perturbation, $J$, and curvature, $K$, for $\delta/\epsilon \gg 1$: $(a,d)$ deep in the Bloch regime, $(b,e)$ near the NIB bifurcation, $(c,f)$ deep in the Ising regime.

disconnected; small $J$ perturbations cannot induce front transitions.

We now demonstrate how an hexagonal advective field, modeling the Bénard convective cells, can induce local front transitions leading to spiral breakup. We numerically integrated (1) with Neumann boundary conditions and $\mathbf{J} = \nabla\phi$ where $\phi = A\sum_{i=1}^{3}\cos(\mathbf{q}_i \cdot \mathbf{x})$ with $\mathbf{q}_1 = (Q,0)$, $\mathbf{q}_2 = (-Q/2, \sqrt{3}Q/2)$, $\mathbf{q}_3 = (-Q/2, -\sqrt{3}Q/2)$. We chose parameters $\epsilon$ and $\delta$ to obtain a $C - J$ relation as depicted in Fig. 3. The upper velocity branch terminates at $J = J_c^+ < 0$, and the lower branch at $J = J_c^- > 0$. The peak value of the advective field, $J_{max} = \text{Max}\{|\mathbf{J}|\}$, was chosen such that

$$|J_c^+| < J_{max} < J_c^-.$$

With this choice of parameters the advective field may induce front transitions from the upper velocity branch to the lower one but not vice versa. The initial conditions, shown in Fig. 4a, pertain to a regular spiral wave in the absence of the advective perturbation. The dark area represents an up state domain. The leading front of the spiral wave (an up state domain invading a down state) and the trailing front (a down state invading an up state) correspond to the upper and lower velocity branches in Fig. 3, respectively.

Figs. 4b, c, d show the time evolution of the initial spiral pattern. As the spiral arm propagates through the hexagonal advective pattern, alternating portions of its leading front experience phases during which $J = \mathbf{J} \cdot \hat{\mathbf{n}} < 0$, i.e. the advective field has a component pointing in

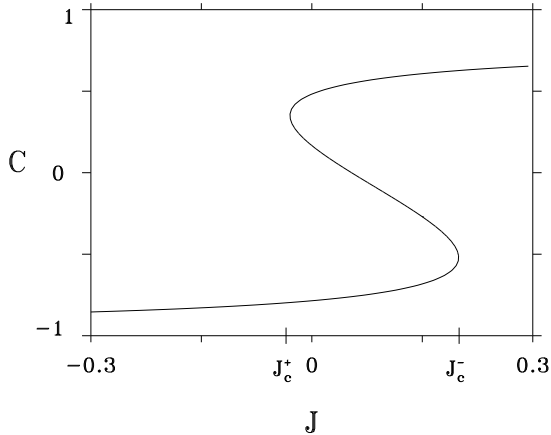

FIG. 3. The relation between the front velocity and the advective field for the parameters used in the spiral breakup simulation shown in Figs. 4 and 5.

a direction opposite to the direction of propagation, $\hat{\mathbf{n}}$. With the above choices of parameter values the advective field is strong enough to drive portions of the leading front from the upper branch, past its end point, to the lower branch. These local transitions nucleate a spiral-vortex pair for each hexagon (Bénard cell) the stripe crosses. These nucleation processes are evident in Figs. 5 (crossings of the $u$ and $v$ zero contour lines) which provide closer looks at the processes occurring in Figs. 4. Portions of the leading front that undergo front transitions eventually annihilate with the trailing front, breaking the stripe into disjoint pieces. Note that the trailing front does not undergo front transitions because the advective field is not strong enough to drive fronts on the lower velocity branch to its end point.

The actual experimental situation is more complex. The system is three dimensional (the flow at the bottom is in the opposite direction to the flow at the top) and we have ignored possible feedback effects of the reaction on the advective field. The mechanism described above can therefore account only for the initial spiral breakup behavior and not for the asymptotic state. We have also considered a bistable medium while the experimental system is excitable, but we expect excitable systems to exhibit similar qualitative behaviors. The front structures that appear in excitable systems coincide with those in bistable systems, to leading order in $\epsilon \ll 1$, and we may expect them to undergo an instability analogous to the NIB bifurcation. Since single front structures do not exist in excitable systems we should rather look at the behavior of front pairs or pulses. Indeed, the NIB bifurcation in bistable systems leads to a pulse instability that has a close analog in excitable systems; depending on the value of $\delta$, traveling pulses in both systems either collapse or develop into breathers as $\epsilon$ is increased past a critical value [3,14,15].

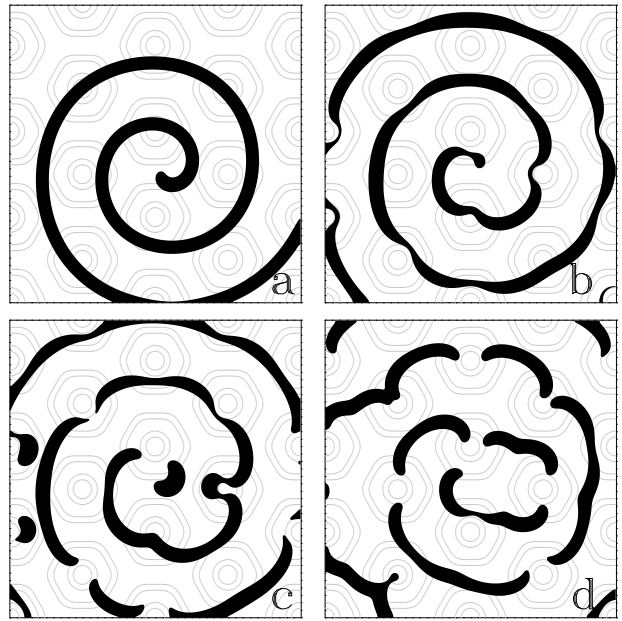

FIG. 4. Breakup of a spiral wave induced by a hexagonal advective pattern. The light and dark regions correspond to down and up states, respectively. The dotted curves denote contours of constant advection speed. The convection flow direction is outward from the centers of the hexagons. Frame $a$ is the unperturbed spiral wave and frames $b, c, d$ are taken at times $t = 100$, $140$, $220$ from the onset of the advective pattern. Parameters used: $a_0 = -0.1$, $a_1 = 2.0$, $\epsilon = 0.032$, $\delta = 0.9$, $A = 1.59$ and $Q = 0.06283$.

## IV. VORTEX NUCLEATION AND DOMAIN SPLITTING INDUCED BY CURVATURE

So far we studied the effects of an extrinsic perturbation in the form of an advective field. Even without extrinsic perturbations there exist intrinsic factors, such as curvature and front interactions, that affect planar front propagation. We now address the effect of curvature near a NIB bifurcation. In Ref. [7] we have found the following relation between the normal velocity of a front, $C$, and its curvature, $K$:

$$C = F(C + \delta K, \eta) - K, \quad (18)$$

where $F$ is given by (14). Graphs of $C$ vs. $K$ deep in the Bloch regime, near a NIB bifurcation, and deep in the Ising regime, are shown in Figs. 2d, 2e and 2f, respectively. A comparison of these figures with Figs. 2a-2c indicates that the multivalued dependence depicted in Fig. 1 is a general feature of front dynamics near a NIB bifurcation. *Perturbations of different nature, curvature and an advective field in this case, can have the same effect of inducing front transitions.*

Recently, an intriguing pattern formation phenomenon has been observed in numerical and laboratory experiments [6,16]. An expanding chemical spot was found to lose its circular shape and split into two spots. Newborn spots followed the same course of events eventually filling the reaction domain. We suggest that spot or domain splitting can result from dynamic front transitions induced by curvature variations. Imagine a $C$ vs. $K$ relation in which the upper branch terminates at a small *positive* curvature value, $K_c^+ > 0$, as depicted in Fig. 6. The leading front of an expanding disk-like domain corresponds to the upper branch (an up state invading a down state). As the domain expands the front curvature decreases. When it falls below, $K_c^+$, the critical curvature pertaining to the edge of the upper branch, a front transition occurs. As a result the domain stops expanding and begins shrinking. We have verified this scenario numerically using a circularly symmetric version of (1) (with $\mathbf{J} = 0$). Two-dimensional realizations of disk-like domains are never perfect; there are always flatter parts of the front which undergo the transition first. Like in the case of advective perturbation, these local front transitions nucleate spiral vortex pairs and lead to domain splitting as shown in Fig. 7 for an initially oval-shaped domain.

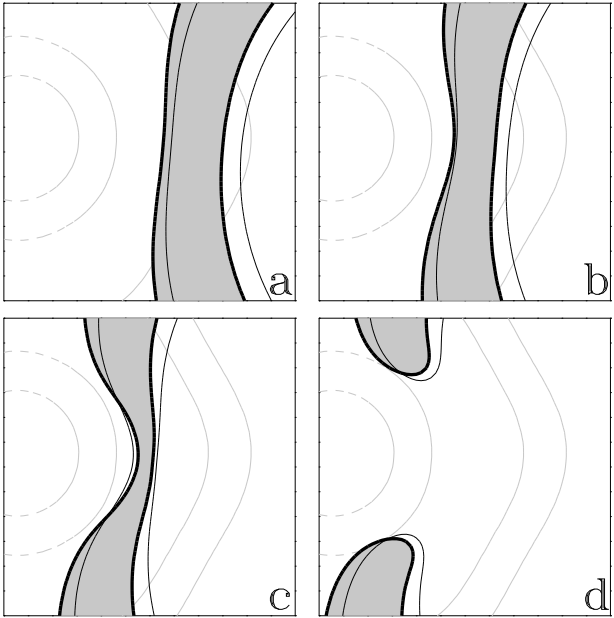

FIG. 5. A closer look at a typical breakup process in Fig. 4. The thick (thin) lines are contours of $u = 0$ ($v = 0$). The direction of front propagation follows from the rule that the $v = 0$ contour always lags behind the $u = 0$ contour. The frames $a, b, c, d$ pertain to times $t = 140, 160, 180, 200$. They show a local front transition, accompanied by the nucleation of a vortex pair (the crossing points of the contour lines), and the breakup of the up state domain. Parameters are the same as in Fig. 4.

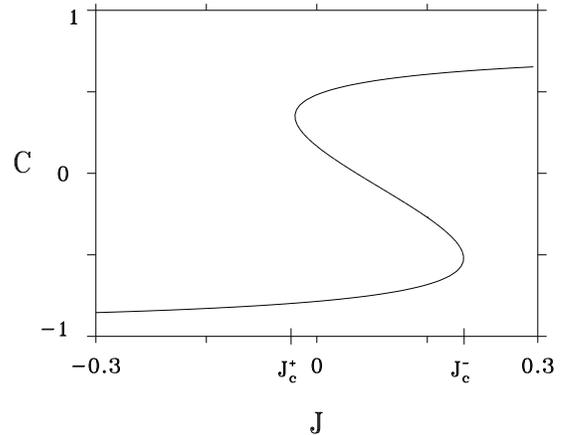

FIG. 6. The relation between front velocity and curvature for the parameters used in the spot splitting simulation shown in Figs. 7.

The transition from labyrinthine patterns to spiral turbulence found in [7] follows from a similar velocity-curvature relation except that the upper branch terminates near zero or at negative curvature values. In that case the mechanism that drives the leading front past the end point is the transverse instability which produces segments with negative curvature.

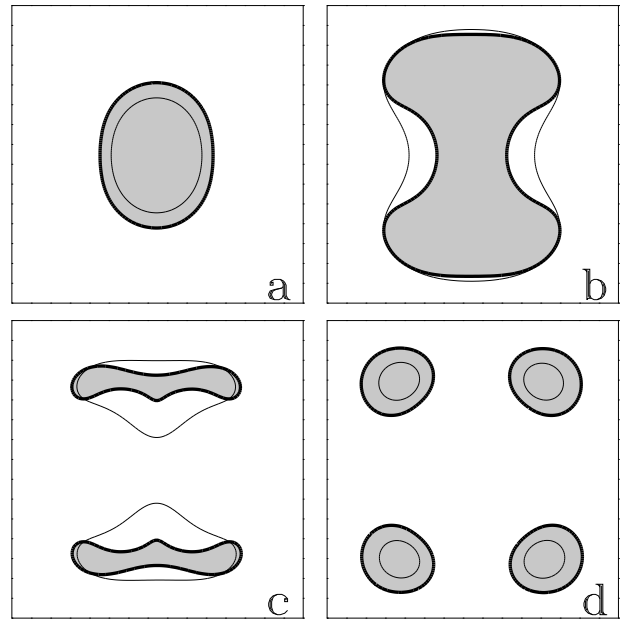

FIG. 7. Spot splitting induced by curvature variations. The frames $a, b, c, d$ pertain to times $t = 80, 240, 280, 340$. Local front transitions occur at the flatter portions of the expanding front. They are accompanied by nucleation of vortex pairs, and followed by spot splitting. Parameters used: $a_0 = -0.15$, $a_1 = 2.0$, $\epsilon = 0.014$, $\delta = 3.5$.

## V. CONCLUSION

We have shown a few examples of local front transitions induced by extrinsic or intrinsic perturbations near a NIB bifurcation. Such transitions are accompanied by the nucleation of spiral vortex pairs, bounding the front segments that underwent the transitions. The spiral vortices can be viewed as "front" structures along the front line where the $v$ field goes from a positive value (pertaining to one Bloch front) to a negative value (pertaining to the other Bloch front), and vice versa. The nucleation of spiral vortices is followed by domain splitting or breakup as Figs. $5c,d$ and $7b,c,d$ show. We note that domain breakup is not a necessary outcome of spiral vortex nucleation. It occurs when the speed at which two fronts approach one another is high enough relative to the rate of diffusive dissipation of $v$ in the region bounded by the fronts. Then, the fronts can approach one another to within a distance of order unity where diffusive dissipation of $u$ becomes effective and can annihilate the fronts. This is usually the case near the NIB bifurcation and beyond it (i.e. in the Bloch regime).

The ideas presented in this paper may explain many more observations. Most recently Taboada *et al.* reported experimental observations, in the BZ reaction, of circular wave breakup phenomena induced by an electric field [8]. They also studied the two-variable Oregonator model with an advective term and numerically reproduced the experimental results. Among their findings is a monotonically increasing relation between the critical electric current amplitude (the advective field) needed for wave breakup and the "excitability", $1/\epsilon$. This finding can be understood once we consider the change in the $C - J$ relation near the NIB bifurcation as $1/\epsilon$ is increased. As Figs. $2a, b, c$ imply, the critical $J$ value, $J_c^+(\epsilon) < 0$, at which the upper velocity branch terminates, increases in absolute value as we go farther into the Bloch regime or increase $1/\epsilon$. In other words, a stronger field $J$ is needed to cause front transitions (and consequently breakup) as the excitability ($1/\epsilon$) is increased.

The onset of breathing motion in pulses near the NIB bifurcation [3,14,17] can also possibly be interpreted in terms of dynamic front transitions. In this case the transitions are induced by intrinsic front interactions that become significant as two fronts approach one another. A multivalued front velocity relation as depicted in Fig. 6 with the horizontal axis replaced by the reciprocal distance between the fronts can account for a breathing motion. Such a relation has not been derived yet.

We have studied an activator-inhibitor type reaction diffusion model but we expect the basic results to be applicable to other systems undergoing a NIB bifurcation. These include periodically forced oscillatory reactions [1] and liquid crystal systems [4].

## ACKNOWLEDGMENTS

We thank Kyoung Lee, John Pearson, Harry Swinney and Art Winfree for stimulating discussions. Most of this work has been done while E. Meron was visiting the Department of Mathematics at the University of Arizona. He wishes to thank Art Winfree, Alan Newell and Jerry Moloney for their wonderful hospitality. A. Hagberg acknowledges the support of the Computational Science Graduate Fellowship Program of the Office of Scientific Computing in the Department of Energy.

* Present address: aric@lanl.gov, Center for Nonlinear Studies, Los Alamos National Laboratory, Los Alamos, NM 87545
[1] P. Coullet, J. Lega, B. Houchmanzadeh, and J. Lajzerowicz, Phys. Rev. Lett. **65**, 1352 (1990).
[2] A. Hagberg and E. Meron, Phys. Rev. E **48**, 705 (1993).
[3] A. Hagberg and E. Meron, Nonlinearity **7**, 805 (1994)
[4] T. Frisch, S. Rica, P. Coullet, and J.M. Gilli, Phys. Rev. Lett. **72**, 1471 (1994); P. Coullet, T. Frisch, J.M. Gilli, and S. Rica, Chaos **4**, 485 (1994).
[5] K.I. Agladze, V.I. Krinsky and A.M. Pertsov, Nature **308**, 834 (1984); S.C. Müller, T. Plesser and B. Hess, in *Physicochemical Hydrodynamics: Interfacial Phenomena*, Ed. M.G. Velarde (Plenum Press, New York, 1988).
[6] K.J. Lee, W.D. McCormick, H.L. Swinney, and J.E. Pearson, Nature **369**, 215 (1994); K.J. Lee and H.L. Swinney, "Lamellar structures and self-replicating spots in a reaction diffusion system", to appear in Phys. Rev. E
[7] A. Hagberg and E. Meron, Phys. Rev. Lett. **72**, 2494 (1994).
[8] J.J. Taboada, A.P. Muñuzuri, V. Pérez-Muñuzuri, M. Gómez-Gesteira, and V. Pérez-Villar, Chaos **4**, 519 (1994).
[9] M. Courtemanche and A.T. Winfree, Int. J. Bifurcation Chaos **1**, 431 (1991); A.V. Holden and A.V. Panfilov, Int. J. Bifurcation Chaos **1**, 219 (1991); A. Karma, Phys. Rev. Lett. **71**, 1103 (1993); M. Bär and M. Eiswirth, Phys. Rev. E **48**, R1635 (1993).
[10] S.P. Dawson, A. Lawniczak, and R. Kapral, J. Chem. Phys. **100**, 5211 (1994).
[11] S. Schmidt and P. Ortoleva, J. Chem. Phys. **67**, 3771 (1977); O. Steinbock, J. Schütze, ans S.C. Müller, Phys. Rev. Lett. **68**, 248 (1992).
[12] A. Hagberg and E. Meron, Chaos **4**, 477 (1994).
[13] H. Ikeda, M. Mimura, and Y. Nishiura, Nonl. Anal. TMA **13**, 507 (1989).
[14] S. Koga and Y. Kuramoto, Prog. Theor. Phys. **63**, 106 (1980).
[15] E. Meron. Phys. Rep. **218**, 1 (1992).
[16] J.E. Pearson, Science **261**, 189 (1993).
[17] Y. Nishiura and M. Mimura, SIAM J. Appl. Math. **49**, 481 (1989).